# High Availability Cluster System for Local Disaster Recovery with Markov Modeling Approach

**T.T.Lwin and T.Thein**

**University of Computer Studies**
**Yangon, Myanmar**

**Abstract**

The need for high availability (HA) and disaster recovery (DR) in IT environment is more stringent than most of the other sectors of enterprises. Many businesses require the availability of business-critical applications 24 hours a day, seven days a week, and can afford no data loss in the event of a disaster. It is vital that the IT infrastructure is resilient with regard to disruption, even site failures, and that business operations can continue without significant impact. As a result, DR has gained great importance in IT. Clustering of multiple industries standard servers together to allow workload sharing and fail-over capabilities is a low cost approach. In this paper, we present the availability model through Semi-Markov Process (SMP) and also analyze the difference in downtime of the SMP model and the approximate Continuous Time Markov Chain (CTMC) model. To acquire system availability, we perform numerical analysis and SHARPE tool evaluation.

## 1. Introduction

High availability clusters (also known as HA Clusters or failover Clusters) are computer clusters implemented to provide high availability of services. They operate by having redundant computers or nodes which are used to provide service when a system component fails.

A cluster is a collection of computer nodes -- independent, self-contained computer systems working together – to provide a more reliable and powerful system than a single node alone [8]. Clustering has proven to be a very effective method for scaling to larger systems for added performance, as well as providing higher levels of availability and lower management costs. For this reason, software packages such as IBM's RS/6000 Cluster Technology [8] (i.e., Phoenix) and Microsoft's Cluster Services [5] (i.e., Wolf pack) are being used to build high availability systems.

Disaster recovery solutions have gained popularity in the past few years because of their ability to tolerate disasters and to achieve the reliability and availability.

## 2. Related Work

Hunter [5] described some system characteristics that benefit from clustering and presented a two-node Microsoft Cluster Service (MSCS) cluster configuration and also presented an availability model of that system using Markov modeling techniques.

In [1] they discussed high availability and disaster recovery solutions, and described how HA and DR solutions differ from one another and how they can be combined to provide the highest levels of resiliency for IT infrastructures.

Trivedi et. al [10] described an availability model for a high availability platform using a multi-level hierarchical composition approach that mixes reliability block diagrams and Markov chains, so as to allow detailed behavior to be captured while avoiding state space explosion.

Song et al [9] provided novel solutions with three –key components, availability modeling, model evaluation and data analysis and examined numerical solutions for Markov models on the uniformization method. This paper also presents a monitoring and data analysis framework, which is responsible for failure analysis and availability reconfiguration.

The semi-Markov decision model is a powerful tool in analyzing sequential decision process with random decision epochs [2]. They presented the application of Markov decision process algorithm, a joint optimization of inspection rate and its corresponding maintenance policy are also presented.

## 3. System Architecture

The architecture is based on an active-passive high availability solution. Each service under high availability needs at least two identical servers: a primary host, on which the service run, one or more secondary hosts, able to recover the application. As a result of failure detection, the active-passive roles are switched. A heartbeat keep-alive system is used to monitor the health of the nodes in the cluster. A disaster recovery solution is typically composed of two nodes, one active and one passive. The active node is usually called master or production node, and the passive node is called secondary or standby node. During normal operation, the only working node is the master node; in the event of a node failover or switchover, the standby node takes over the production role, by taking its IP number, and completely replacing the master one.

To maintain the standby node for failover, the standby node contains homogenous installations and applications: data and configurations must also be constantly synchronized with the master node.

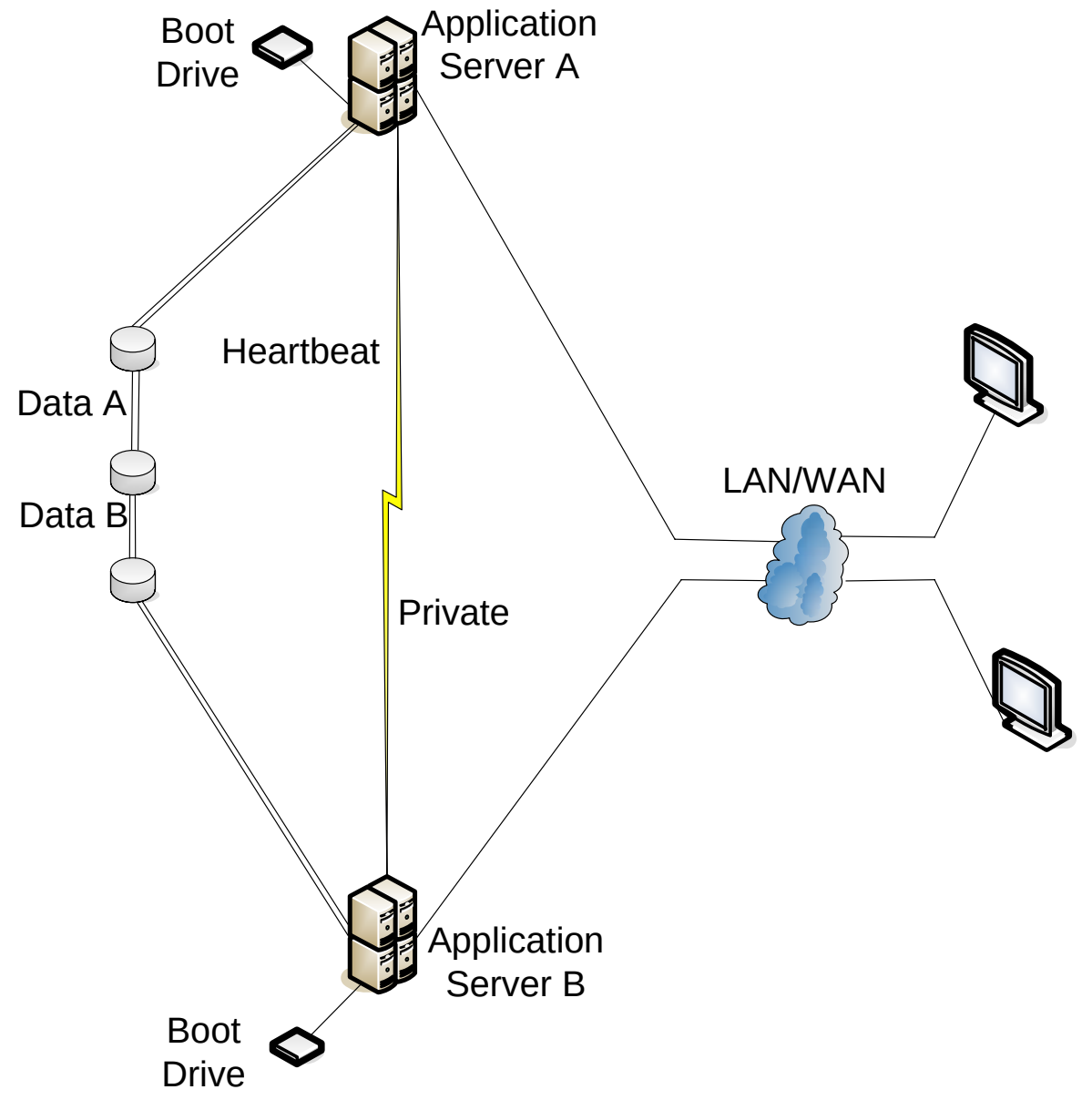

Figure1: System Framework

If a crash occurs and if the data is not restored, it can have devastating consequences for a business. So it is imperative for companies to effectively backup and recover data and protect them from huge losses in productivity and downtime.

In this way, hardware exposure is mitigated through physical hardware redundancy. Clustering provides high availability by protecting against a node failure. However, it does not prevent against storage failures. Given the size of typical cluster environments, multiple hard disks are used to build large storage arrays. In Network and System Administration, when large numbers of any one device are used, failure is expected. When a hard disk fails, application disruption is unavoidable, as all the nodes in the cluster could be using that one particular disk as shared storage which contains all files.

With the widespread use of computers, data is becoming more and more important in human life. But all kinds of accidents and disasters occur frequently. Data corruption and data loss by various disasters have become more dominant, accounting for over 60% [1] of data loss .Recent high-profile data loss has raised awareness of the need to plan for recovery of continuity. Many data disaster tolerance technologies have been employed to increase the availability of data and to reduce the data damage caused by disasters [2].

A true disaster recovery solution is the ability to restore full systems quickly on available computing resources which may be local but may also be remote if the situation dictates and must allow recovery from site-wide disasters. The primary site may be completely down, a secondary site located in a non-affected area would be used to restore services until the primary site comes back online.

## 4. Modeling and Analysis

We propose the two-component system, one component is considered as active and the other as a standby (spare) unit. The failure rates of the active unit and the standby unit are different, and also the effect of failure of the standby unit is different from that of the active unit. Assuming that, the time to restoration and reboot are exponentially distributed with rate $\mu$ and $\beta$ respectively.

We consider a routine diagnostic that is run every T time units, intended to detect the latent fault of the standby unit. While units' failure and restoration times are exponentially distributed, the routine diagnostic time interval is not a continuous time Markov chain. The model for the system with the diagnostic routine is called a semi-Markov chain. To solve this model, we could crudely approximate the time to the next diagnostic to be exponentially distributed with mean $\frac{T}{2}$ .Descriptions of the state are shown in table (1).

Table (1): State Description for Transitions model

| State | Descriptions |
|---|---|
| 1 | Both active and spare units are working |
| 2 | Protection switch fails to cover the failure of the active unit |
| 3 | When active unit fails, protection switch successfully restores service by the standby unit |
| 4 | The failure of the standby unit while the active unit is still working is detected immediately |
| 5 | The failure of the standby unit is not detected |
| 6 | The system is in failure state |

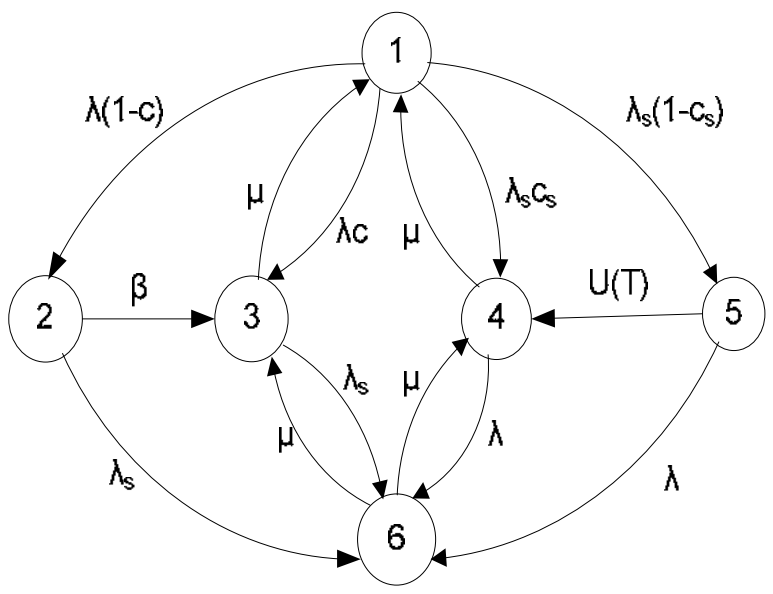

Figure2 : State Transition Model

$\lambda$=failure rate of an active unit
$\lambda_s$=failure rate of a standby unit
$\mu$=restoration rate of a failed unit
c =coverage probability of an active unit
$c_s$=coverage probability of a standby unit
T =time units to detect the latent fault of the standby unit

We may compute the steady-state probabilities by first writing down the steady-state balance equations of figure 2 are as follows:

$$\mu P_3 + \mu P_4 = \lambda(1-c)P_1 + \lambda c P_1 + \lambda_s c_s P_1 + \lambda_s(1-c_s)P_1 \quad (1)$$

$$\lambda(1-c)P_1 = (\beta + \lambda_s)P_2 \quad (2)$$

$$\lambda c P_1 + \beta P_2 + \mu P_6 = \mu P_3 + \lambda_s P_3 \quad (3)$$

$$\lambda_s c_s P_1 + \frac{2}{T}P_5 + \mu P_6 = (\mu + \lambda)P_4 \quad (4)$$

$$\lambda_s(1-c_s)P_1 = \left(\frac{2}{T} + \lambda\right)P_5 \quad (5)$$

$$\lambda_s P_2 + \lambda_s P_3 + \lambda P_4 + \lambda P_5 = 2\mu P_6 \quad (6)$$

The conservation equation of figure 2 is obtained by summing the probabilities of all states in the system and the sum of the equation is 1.

$$\sum_{i=1}^{n} P_i = 1 \quad (7)$$

Combining the above-mentioned balance equations with the conservation equations, and solving these simultaneous equations, we acquire the closed-form solution for the system.

$$P_1 = \left[ 1 + \frac{\lambda(1-c)}{\lambda_s+\beta} - \frac{1}{\mu+\lambda_s}\left(\lambda_s c_s + \frac{\lambda_s(1-c_s)\frac{2}{T}}{\frac{2}{T}+\lambda} - \lambda c - \frac{\lambda(1-c)\beta}{\lambda_s+\beta}\right) + \frac{\lambda_s(1-c_s)}{\frac{2}{T}+\lambda} - \frac{\lambda_s c_s}{\mu} - \frac{\lambda_s(1-c_s)\frac{2}{T}}{\mu\left(\frac{2}{T}+\lambda\right)} + \left(\frac{\mu+\lambda}{\mu+\lambda_s} + 1 + \frac{\mu+\lambda}{\mu}\right)\left(\frac{\lambda(1-c)+\lambda c+\lambda_s c_s+\lambda_s(1-c_s) + \frac{\mu}{\mu+\lambda_s}\left(\lambda_s c_s + \frac{\lambda_s(1-c_s)\frac{2}{T}}{\frac{2}{T}+\lambda} - \lambda c - \frac{\lambda(1-c)\beta}{\lambda_s+\beta}\right)}{\frac{\mu(\mu+\lambda)}{\mu+\lambda_s} + \mu}\right) \right]^{-1} \quad (8)$$

$$P_2 = \left(\frac{\lambda(1-c)}{\lambda_s+\beta}\right)P_1 \quad (9)$$

$$P_3=\left(\begin{array}{l}\frac{\mu+\lambda}{\mu+\lambda_s}\times\\ \lambda(1-c)+\lambda c+\lambda_s c_s+\lambda_s(1-c_s)+\\ \frac{\mu}{\mu+\lambda_s}\\ \frac{\left(\lambda_s c_s+\frac{\lambda_s(1-c_s)\frac{2}{T}}{\frac{2}{T}+\lambda}-\lambda c-\frac{\lambda(1-c)\beta}{\lambda_s+\beta}\right)}{\frac{\mu(\mu+\lambda)}{\mu+\lambda_s}+\mu}\\ -\frac{1}{\mu+\lambda_s}\\ \left(\lambda_s c_s+\frac{\lambda_s(1-c_s)\frac{2}{T}}{\frac{2}{T}+\lambda}-\lambda c-\frac{\lambda(1-c)\beta}{\lambda_s+\beta}\right)\end{array}\right)P_1 \quad (10)$$

$$P_4=\left(\begin{array}{l}\lambda(1-c)+\lambda c+\lambda_s c_s+\lambda_s(1-c_s)+\\ \frac{\mu}{\mu+\lambda_s}\\ \frac{\left(\lambda_s c_s+\frac{\lambda_s(1-c_s)\frac{2}{T}}{\frac{2}{T}+\lambda}-\lambda c-\frac{\lambda(1-c)\beta}{\lambda_s+\beta}\right)}{\frac{\mu(\mu+\lambda)}{\mu+\lambda_s}+\mu}\end{array}\right)P_1 \quad (11)$$

$$P_5=\left(\frac{\lambda_s(1-c_s)}{\frac{2}{T}+\lambda}\right)P_1 \quad (12)$$

$$P_6=\left(\begin{array}{l}\frac{\mu+\lambda}{\mu}\times\\ \lambda(1-c)+\lambda c+\lambda_s c_s+\lambda_s(1-c_s)+\\ \frac{\mu}{\mu+\lambda_s}\\ \frac{\left(\lambda_s c_s+\frac{\lambda_s(1-c_s)\frac{2}{T}}{\frac{2}{T}+\lambda}-\lambda c-\frac{\lambda(1-c)\beta}{\lambda_s+\beta}\right)}{\frac{\mu(\mu+\lambda)}{\mu+\lambda_s}+\mu}\\ -\frac{\lambda_s c_s}{\mu}-\frac{\lambda_s(1-c_s)\frac{2}{T}}{\mu\left(\frac{2}{T}+\lambda\right)}\end{array}\right)P_1 \quad (13)$$

## 4.1Semi-Markov Model Analysis

A better approach would be to take the time the next diagnostic to be uniformly distributed over [0, T], resulting in a semi-Markov chain. This is indicated in fig: 2 the transition labeled U (0, T). As occurring in two stages of transitions, the SMP is described by a transition probability matrix P and the vector of sojourn time distributions, **H** (t).

$$H_1=1-e^{-(\lambda+\lambda_s)t} \quad (14)$$

$$H_2=1-e^{-(\beta+\lambda_s)t} \quad (15)$$

$$H_3=1-e^{-(\lambda_s+\mu)t} \quad (16)$$

$$H_4=1-e^{-(\lambda+\mu)t} \quad (17)$$

$$H_5=\begin{cases}1-\left(1-\frac{t}{T}\right)e^{-\lambda t}, t<T,\\ 1, t\geq T,\end{cases} \quad (18)$$

$$H_6=1-e^{2\mu t} \quad (19)$$

Let X~EXP (λ) and Y~U (0, T) random variables

$$P(X>Y)=\int_0^T P(X>t)f_Y(t)dt$$

$$=\int_0^T e^{-\lambda t}\frac{1}{T}dt=\frac{1}{\lambda T}\left(1-e^{-\lambda T}\right) \qquad (20)$$

The one-step transition probability matrix P of the DTMC embedded at the time of transitions and the state probabilities of the embedded DTMC are given by the following equations respectively.

$$P=\begin{array}{c} \\ 1\\2\\3\\4\\5\\6\end{array}\begin{array}{c}\begin{array}{cccccc}1&2&3&4&5&6\end{array}\\ \begin{bmatrix} 0 & \frac{\lambda(1-c)}{\lambda+\lambda_s} & \frac{\lambda c}{\lambda+\lambda_s} & \frac{\lambda_s c_s}{\lambda+\lambda_s} & \frac{\lambda_s(1-c_s)}{\lambda+\lambda_s} & 0 \\ 0 & 0 & \frac{\beta}{\beta+\lambda_s} & 0 & 0 & \frac{\lambda_s}{\beta+\lambda_s} \\ \frac{\mu}{\lambda_s+\mu} & 0 & 0 & 0 & 0 & \frac{\lambda_s}{\lambda_s+\mu} \\ \frac{\mu}{\lambda+\mu} & 0 & 0 & 0 & 0 & \frac{\lambda}{\lambda+\mu} \\ 0 & 0 & 0 & \frac{1}{\lambda T\left(1-e^{-\lambda T}\right)} & 0 & 1-\frac{1}{\lambda T}\left(1-e^{-\lambda T}\right) \\ 0 & 0 & \frac{1}{2} & \frac{1}{2} & 0 & 0 \end{bmatrix}\end{array} \qquad (21)$$

$$v=\left[v_{1,1},v_{1C},v_{0,1},v_{1D},v_{0,0}\right] \qquad (22)$$

To obtain the steady state probabilities, solve the equation

**v=vP** (23)

This yield

$$v_2=\frac{\lambda(1-c)}{\lambda+\lambda_s}v_1 \qquad (24)$$

$$v_3=\frac{\lambda_s+\mu}{\mu}\left[1-\frac{\mu(\lambda_s+\mu)}{\lambda\mu+\lambda_s\mu+2\mu^2}\left(1+\frac{1}{\lambda+\lambda_s}\left(\lambda_s c_s+\frac{1}{\lambda T}\left(1-e^{-\lambda T}\right)\lambda_s(1-c_s)-\lambda c-\frac{\lambda(1-c)\beta}{\beta+\lambda_s}\right)\right)\right]v_1 \qquad (25)$$

$$v_4=\left[\frac{(\lambda_s+\mu)(\lambda+\mu)}{\lambda\mu+\lambda_s\mu+2\mu^2}\left(1+\frac{1}{\lambda+\lambda_s}\left(\lambda_s c_s+\frac{1}{\lambda T}\left(1-e^{-\lambda T}\right)\lambda_s(1-c_s)-\lambda c-\frac{\lambda(1-c)\beta}{\beta+\lambda_s}\right)\right)\right]v_1 \qquad (26)$$

$$v_5=\frac{\lambda_s(1-c_s)}{\lambda+\lambda_s}v_1 \qquad (27)$$

$$v_6=2\left[\frac{\lambda_s+\mu}{\mu}-\frac{(\lambda_s+\mu)^2}{\lambda\mu+\lambda_s\mu+2\mu^2}\left(1+\frac{1}{\lambda+\lambda_s}\left(\lambda_s c_s+\frac{1}{\lambda T}\left(1-e^{-\lambda T}\right)\lambda_s(1-c_s)-\lambda c-\frac{\lambda(1-c)\beta}{\beta+\lambda_s}\right)\right)-\frac{\lambda c}{\lambda+\lambda_s}-\frac{\lambda(1-c)\beta}{(\lambda+\lambda_s)(\beta+\lambda_s)}\right]v_1 \qquad (28)$$

The mean sojourn time $h_i$ in state i is

$$h_i = \int_0^{\infty} \left(1 - H_i(t)\right) dt \tag{29}$$

$$h_1 = \frac{1}{\lambda + \lambda_s} \tag{30}$$

$$h_2 = \frac{1}{\beta + \lambda_s} \tag{31}$$

$$h_3 = \frac{1}{\lambda_s + \mu} \tag{32}$$

$$h_4 = \frac{1}{\lambda + \mu} \tag{33}$$

$$h_5 = \frac{1}{\lambda} - \frac{1}{T\lambda^2}\left(1 - e^{-\lambda T}\right) \tag{34}$$

$$h_6 = \frac{1}{2\mu} \tag{35}$$

The state probabilities of the semi-Markov chain are

$$\pi_i = \frac{v_i h_i}{\sum_j v_j h_j} \tag{36}$$

, where $i, j \in \{(1,1), 1C, (0,1), (1,0), 1D, (0,0)\}$

$$\pi_1 = \frac{h_1}{h_1 + \frac{\lambda(1-c)}{\lambda+\lambda_s} h_2 + \left[\frac{\lambda_s+\mu}{\mu} - \frac{(\lambda_s+\mu)^2}{\lambda\mu+\lambda_s\mu+2\mu^2}\left(1 - \frac{1}{\lambda+\lambda_s}\left(\lambda_s c_s + \frac{1}{\lambda T}\left(1-e^{-\lambda T}\right)\lambda_s(1-c_s) - \lambda c - \frac{\lambda(1-c)\beta}{\beta+\lambda_s}\right)\right)\right] h_3 + \left[\frac{(\lambda_s+\mu)(\lambda+\mu)}{\lambda\mu+\lambda_s\mu+2\mu^2}\left(1 + \frac{1}{\lambda+\lambda_s}\left(\lambda_s c_s + \frac{1}{\lambda T}\left(1-e^{-\lambda T}\right)\lambda_s(1-c_s) - \lambda c - \frac{\lambda(1-c)\beta}{\beta+\lambda_s}\right)\right)\right] h_4 + \frac{\lambda_s(1-c_s)}{\lambda+\lambda_s} h_5 + \left[\frac{2(\lambda_s+\mu)}{\mu} - \frac{2(\lambda_s+\mu)^2}{\lambda\mu+\lambda_s\mu+2\mu^2}\left(1 + \frac{1}{\lambda+\lambda_s}\left(\lambda_s c_s + \frac{1}{\lambda T}\left(1-e^{-\lambda T}\right)\lambda_s(1-c_s) - \lambda c - \frac{\lambda(1-c)\beta}{\beta+\lambda_s}\right)\right) - \frac{2\lambda c}{\lambda+\lambda_s} - \frac{2\lambda(1-c)\beta}{(\lambda+\lambda_s)(\beta+\lambda_s)}\right] h_6} \tag{37}$$

$$\pi_2 = \frac{\lambda(1-c)}{\lambda+\lambda_s} h_2 \times \frac{P_1}{h_1} \tag{38}$$

$$\pi_3 = \left[\frac{\lambda_s+\mu}{\mu} - \frac{(\lambda_s+\mu)^2}{\lambda\mu+\lambda_s\mu+2\mu^2}\left(1 - \frac{1}{\lambda+\lambda_s}\left(\lambda_s c_s + \frac{1}{\lambda T}\left(1-e^{-\lambda T}\right)\lambda_s(1-c_s) - \lambda c - \frac{\lambda(1-c)\beta}{\beta+\lambda_s}\right)\right)\right] h_3 \times \frac{\pi_1}{h_1} \tag{39}$$

$$\pi_4 = \left[ \frac{(\lambda_s+\mu)(\lambda+\mu)}{\lambda\mu+\lambda_s\mu+2\mu^2} \left( 1+\frac{1}{\lambda+\lambda_s} \left( \lambda_s c_s + \frac{1}{\lambda T}\left(1-e^{-\lambda T}\right) \lambda_s(1-c_s) - \lambda c - \frac{\lambda(1-c)\beta}{\beta+\lambda_s} \right) \right) \right] \quad (40)$$

$$\pi_5 = \frac{\lambda_s(1-c)}{\lambda+\lambda_s} h_5 \times \frac{P_1}{h_1} \quad (41)$$

$$\pi_6 = \left[ \frac{\lambda_s+\mu}{\mu} - \frac{(\lambda_s+\mu)^2}{\lambda\mu+\lambda_s\mu+2\mu^2} \left( 1-\frac{1}{\lambda+\lambda_s} \left( \lambda_s c_s + \frac{1}{\lambda T}\left(1-e^{-\lambda T}\right)\lambda_s(1-c_s) - \lambda c - \frac{\lambda(1-c)\beta}{\beta+\lambda_s} \right) \right) - \frac{2\lambda c}{\lambda+\lambda_s} - \frac{2\lambda(1-c)\beta}{(\lambda+\lambda_s)(\beta+\lambda_s)} \right] h_6 \times \frac{P_1}{h_1} \quad (42)$$

## 5. Experimental Results

The exact model parameter values for the model are not known, however, a good estimate value for a range of model parameter is assumed. Fig: 3 plots the difference between downtime (minutes per year) estimates obtained using the SMP model and that obtained by approximating the U (0, T) distribution by an exponential distribution with mean T/2. We take the values c=0.9, $c_s$=0.9, μ=1per hour, β=12 per hour, and $\lambda_s$= λ/4. We see that the higher the μ/λ ratio, the lower the downtime computed by the two models.

Availability models capture failure and repair behavior of systems and their components. States of the underlying Markov chain will be classified as up states or down states. The system is not available in the state 2 and state 6. The system availability in the steady-state is defined as follows:

Availability=1-Unavailability
=1-($\pi_2$+$\pi_6$) (43)

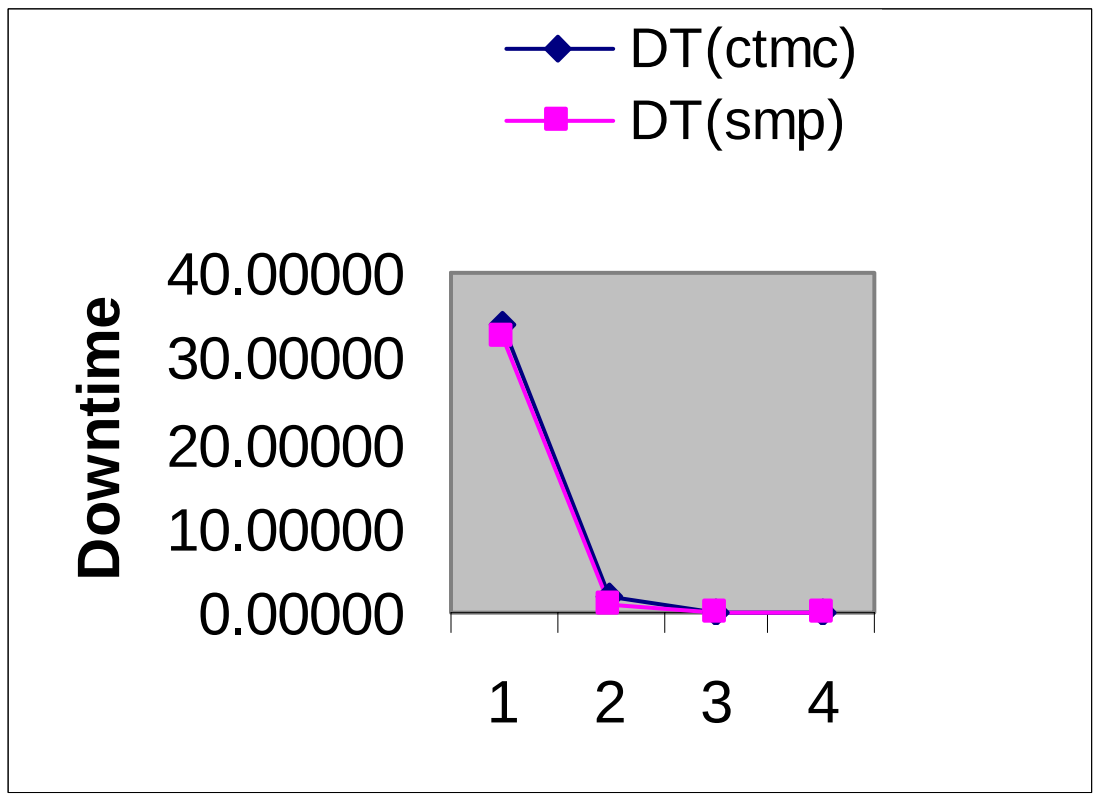

Figure3:Difference in downtime of the SMP model and the approximate CTMC model

### 5.1 Validation of Closed-form Results

To verify the validity of our formula derivations, we compare the results obtained from the closed-form solution and the results obtained from the numerical solution by SHARPE. We found that our results are same.

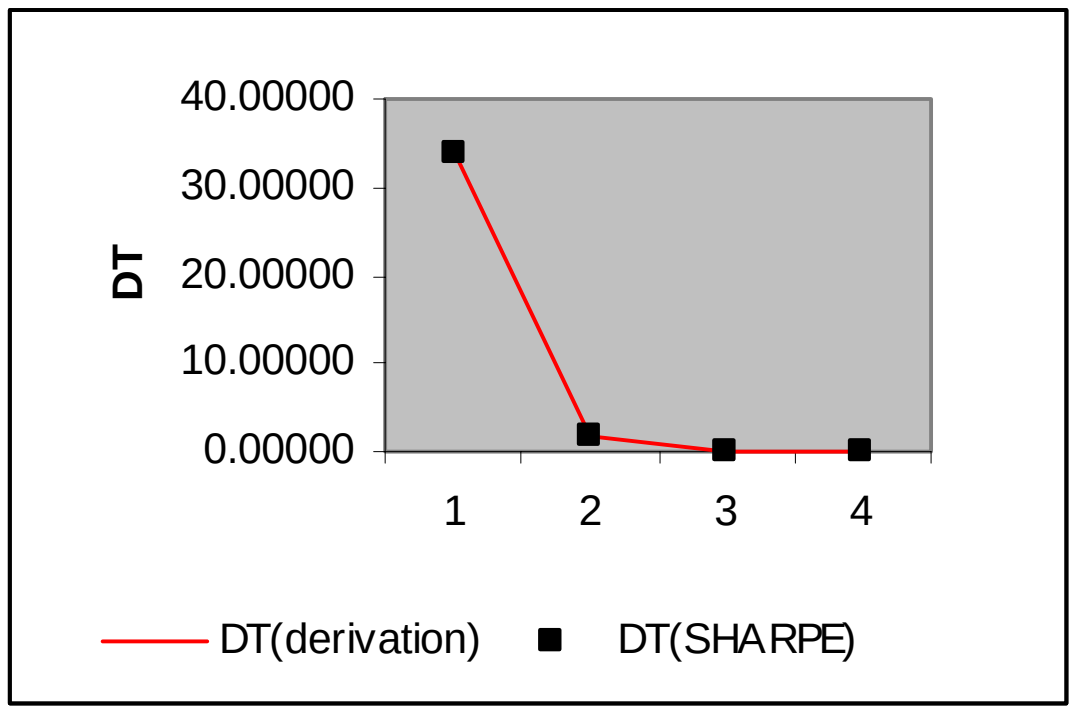

Figure4: Downtime of the CTMC model

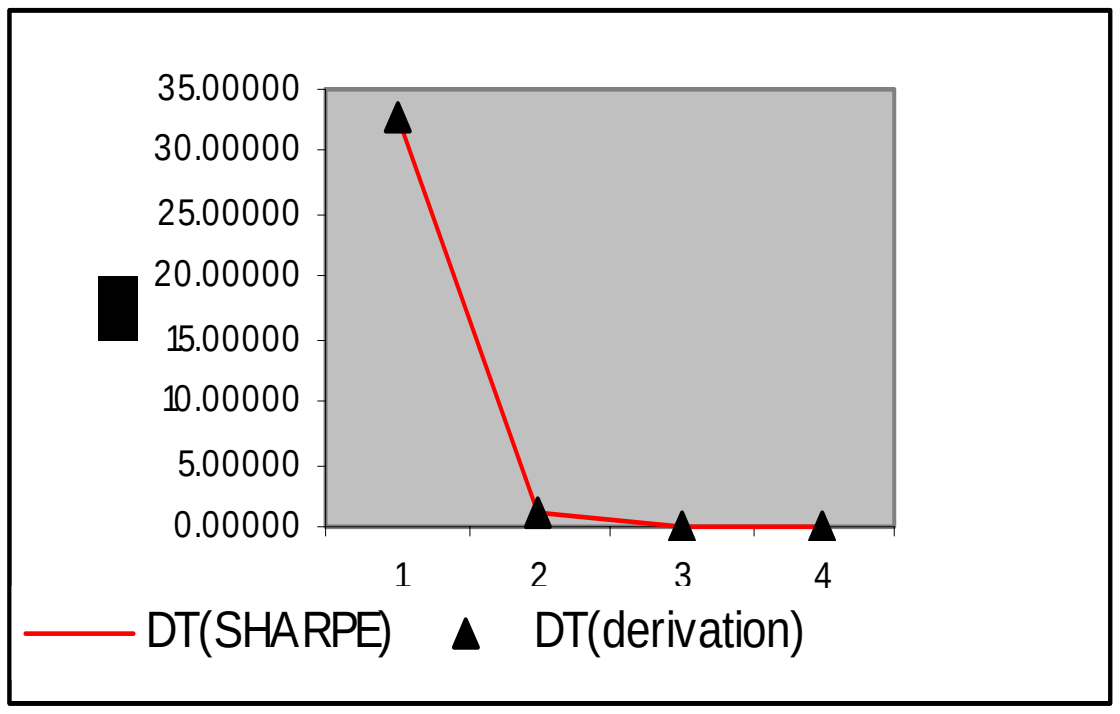

Figure5: Downtime of the SMP model

## 6. Conclusion

Organizations today face a tough challenge in choosing an appropriate high availability solution that meets their business requirements and IT budgets. To implement this requirement, organizations must give high availability and disaster recovery. High availability systems require fewer failures and faster repair. In this paper we presented high availability cluster and failover availability for disaster events. . We present a Markov model and express availability and downtime in terms of the parameters in the model. We evaluate the feasibility of our clustering model using SHARPE tools.

## References

[1] D. Clitherow, M. Brookbanks, N. Clayton, and G. Spear, ‘‘Combining High Availability and Disaster Recovery Solutions for Critical IT Environments,’’ IBM Systems Journal 47, No. 4, 563–575 (2008)

[2] D.Chen, K.S.Trivedi,”Optimization for condition-based maintenance with semi-Markov decision process” Available online at www.sciencedirect.com

[3] R. Gamache, R Short, and Mike Massa, "Windows NT Clustering Service," IEEE Computer, October 1998, pp.55-61.

[4] C.Hirel, A. Robin, Sahner, X.Zang, K.S.Trivedi: “Reliability and performing modeling using SHARPE 2000”. Computer Performance Evaluation/TOOLS 2000. In Lecture Notes in Computer Science; Vol.1786, Springer-Verlag, 2000, pp.345-349.

[5] S. W. Hunter and W. E. Smith, “Availability Modeling and Analysis of a Two Node Cluster,” Proceedings of the 5th International Conference on Information Systems, Analysis and Synthesis, Orlando, FL, October 1999.

[6] Th. Lumpp, J. Schneider, J. Holtz, M. Mueller, N. Lenz, A. Biazetti, and D. Petersen, ‘‘From High Availability and Disaster Recovery to Business Continuity Solutions,’’ IBM Systems Journal 47, No. 4, 605–619

[7] M.Malhotra, A.Reibman:”Selecting and Implementing Phase Approximations for Semi-Markov Models”, Volume 9, Issue 4, 1993, Pages 473-506.

[8] G.F. Pfister, In Search of Clusters: The Coming Battle in Lowly Parallel Computing, Prentice Hall, Englewood Cliffs, NJ, 1998.

[9] H.Song, C.Leangsuksun, R.Nassar, “Availability Modeling and Evaluation on High Performance Cluster Computing Systems,” Journal of Research and Practice in Information Technology, Vol.38, No.4, November 2006.

[10] K. S. Trivedi, R. Vasireddy, D. Trindade, S. Nathan, and R. Castro. Modeling high availability systems. In Proc. Pacific Rim Dependability Conference, 2006.

[11] K.S.Trivedi:” Probability and Statistics with Reliability, Queuing, and Computer Science Applications”, John Wiley and Sons, 2002.

[12] M.Wiboonrat, “Transformation of System Failure Life Cycle,” International Journal of Management Science and Engineering Management, Vol.4 (2008) No.2, pp.143-152.